\documentclass[dvips]{article}

\usepackage{icrc2011}

\title{Highlights of the MAGIC Telescopes}

\shorttitle{J. Cortina for MAGIC coll., MAGIC Highlights}

\authors{Juan Cortina$^{1}$, for the MAGIC Collaboration}
\afiliations{$^1$Institut de Fisica d'Altes Energies, Barcelona, Spain}
\email{cortina@ifae.es}

\abstract{
The MAGIC two 17 meter diameter Very High Energy (VHE) $\gamma$-ray telescopes have now operated 
for two years in stereoscopic mode. The performance of the instrument has been evaluated:
the integral sensitivity for an energy above 300~GeV is 0.76\% crab units
(10\% Crab units differential sensitivity below 100~GeV) and the analysis threshold 
energy is 50 GeV. Highlights of the last two years of observations are the
measurement of the Crab Nebula spectrum from $\sim$50~GeV to $\sim$50~TeV;
the detection of the Crab pulsar up to an energy of 400~GeV, with energy spectra measured 
for both P1 and P2; the discovery of two new radiogalaxies at VHE (NGC~1275 and IC-310);
the absence of an energy cutoff and the discovery of fast variability in the quasars 3C~279 
and PKS~1222+21; the discovery at VHE and the characterization of numerous blazars; 
upper limits to the VHE emission of the Perseus cluster of galaxies and to Dark Matter 
annihilation in dwarf Spheroidals and the measurement of the electron+positron spectrum 
between 100~GeV and 3~TeV. MAGIC is currently undergoing a major upgrade of the readout 
and trigger electronics, and of the camera of the first telescope.}
\keywords{ 
Highlights, MAGIC telescopes, Very High Energy gamma rays, VHE, IACT, pulsars, 
galactic, extragalactic, clusters of galaxies, dark matter. }

\begin{document}
\maketitle

\section{MAGIC, performance of the stereo system}

The two MAGIC Imaging Atmospheric Cherenkov telescopes (IACTs)
were built and are currently operated by a collaboration of
institutions in Bulgaria (Sofia), Croatia (Croatian consortium), Finland (Tuorla Observatory),
Germany (DESY Zeuthen, Universities of Dortmund and W\"urzburg and MPI for Physics in Munich),
Italy (INFN Padova, INFN Siena/Pisa, University of Siena, INFN Udine, an INAF consortium and 
University of Insubria), Poland (University of Lodz), Spain (Universities of Barcelona, Aut\'onoma Barcelona and 
Complutense Madrid, and IEEC-CSIC, IFAE, IAA and IAC), and Switzerland (ETH Zurich)
\footnote{An updated list of collaboration members can be found at {\it http://magic.mppmu.mpg.de}}.

\begin{figure}[!t]
  \vspace{5mm} \centering
  \includegraphics[width=0.95\columnwidth]{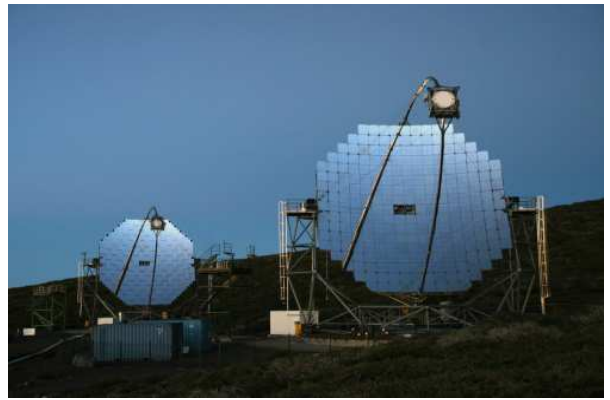}
  \caption{\small{A picture of the two MAGIC telescopes at the Roque de los Muchachos observatory.}}
  \label{fig:magic}
\end{figure}

MAGIC is located at the Roque de los Muchachos Observatory in the Canarian island La Palma
(Spain) at 28$^\circ$46' N, 17$^\circ$53' W and a height of 2200 m.a.s.l. 
The first single telescope (MAGIC-I) remains the largest IACT yet constructed (17m diameter mirror),
a fact which translates into the lowest energy threshold for VHE $\gamma$-ray detection. 
MAGIC-I featured significant novelties in IACTs, such as the fastest
sampling of Cherenkov signals to date (2 GSps) or active mirror control. Its ultralight carbon fiber 
frame and mirrors enable very fast repositioning of the telescope ($<$20 secs for half a turn),
a crucial fact to study the prompt emission of GRBs.

\begin{figure}[!t]
  \vspace{5mm} \centering
  \includegraphics[width=0.95\columnwidth]{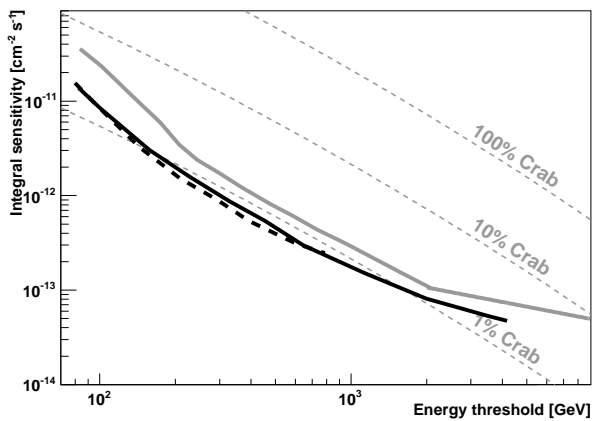}
  \caption{\small{
      Integral sensitivity of the MAGIC Stereo system, defined as
      the flux above a given energy for which $N_{\rm
        excess}/\sqrt{N_{\rm bgd}}=5$ after 50~h. The solid
      black line shows the values obtained from the Crab data. The
      dashed black line shows the values obtained from MC estimations (MC
      gamma rays and MC backgrounds including protons, helium and
      electrons). In the same figure the MAGIC~I sensitivity is shown as
      a grey solid line. Different fractions of the Crab
      Nebula flux (dashed grey) are also shown. From \cite{performance}.
    }}
  \label{fig:sensitivity}
\end{figure}

The introduction of a second telescope, MAGIC-II, enabled the instrument to perform stereoscopic
observations with significantly better sensitivity, and angular and spectral resolutions.
Regular observations with the two telecopes started in Fall 2009. See Fig. \ref{fig:magic}
for a picture of the two IACT system. MAGIC-II was built essentially as
an improved copy of MAGIC-I. The main difference between the telescopes are
their cameras. MAGIC~I camera is composed of 577 hexagonal pixels with
an angular size of 0.1$^\circ$ in the inner part of the camera and
0.2$^\circ$ in the outer part. On the contrary, the MAGIC~II camera is composed
of 1039 0.1$^\circ$ hexagonal pixels.

The threshold energy (peak of the energy distribution of stereo recorded events) has been estimated 
to be 50 GeV assuming a differential spectral index of 2.6. 
Fig. \ref{fig:sensitivity} shows the integral sensitivity of the instrument as a function of 
energy. For energy above 300~GeV the sensitivity is 0.76\% Crab units. There is 
a good agreement with the predictions from MC simulations. Respect to single telescope
(``mono'') observations, a factor $\sim$2 improvement in significance is achieved at a 
few hundred GeV and up to a factor $\sim$3 at lower energies. The differential sensitivity
remains acceptable (10\% Crab units) below 100~GeV.  

Stereo observations result in a significant improvement both in angular and spectral resolutions.
An angular resolution of 0.07$^\circ$ is reached at 300~GeV. The best spectral resolution
of 16\% is reached at a few hundred GeV. Find more details about the instrument's performance
in \cite{icrc_performance,performance}.

\section{Galactic observations}

\subsection{Crab Nebula}

The Crab Nebula is the standard candle of VHE astronomy and an archetypal Pulsar Wind Nebula.
At VHE the spectrum is dominated by Inverse Compton emission. 
Crucial for understanding this emission is the determination of the peak energy and the 
clarification of a possible spectrum cutoff at energies above 
10~TeV due to Klein-Nishina effects. MAGIC has observed Crab for 49~hours during the past 
two winter seasons (2009/10 and 2010/11\cite{icrc_Crab_nebula}) at zenith angles between 
0$^\circ$ and 50$^\circ$. New analysis techniques have been employed to
achieve the lowest possible threshold (see \cite{icrc_advanced_tech}). This large data sample and
the improved sensitivity of the instrument permit producing a spectrum spanning three orders of magnitud in
energy from $\sim$50~GeV to $\sim$50~TeV, with six spectral points for every decade of energy
and first spectral point at 58 GeV. The spectrum can be fitted to a ``curved power law'':
\\
\\
$\frac{dN}{dE} [TeV^{-1} cm^{-2} s^{-1}] = $
\\
$ (3.20\pm 0.05)~10^{-11} (\frac{E}{TeV})^{ (-2.4\pm 0.1) + (0.13\pm 0.02) log (\frac{E}{TeV})} $
\\
\\

\begin{figure}
  \centering
  \includegraphics[width=\columnwidth]{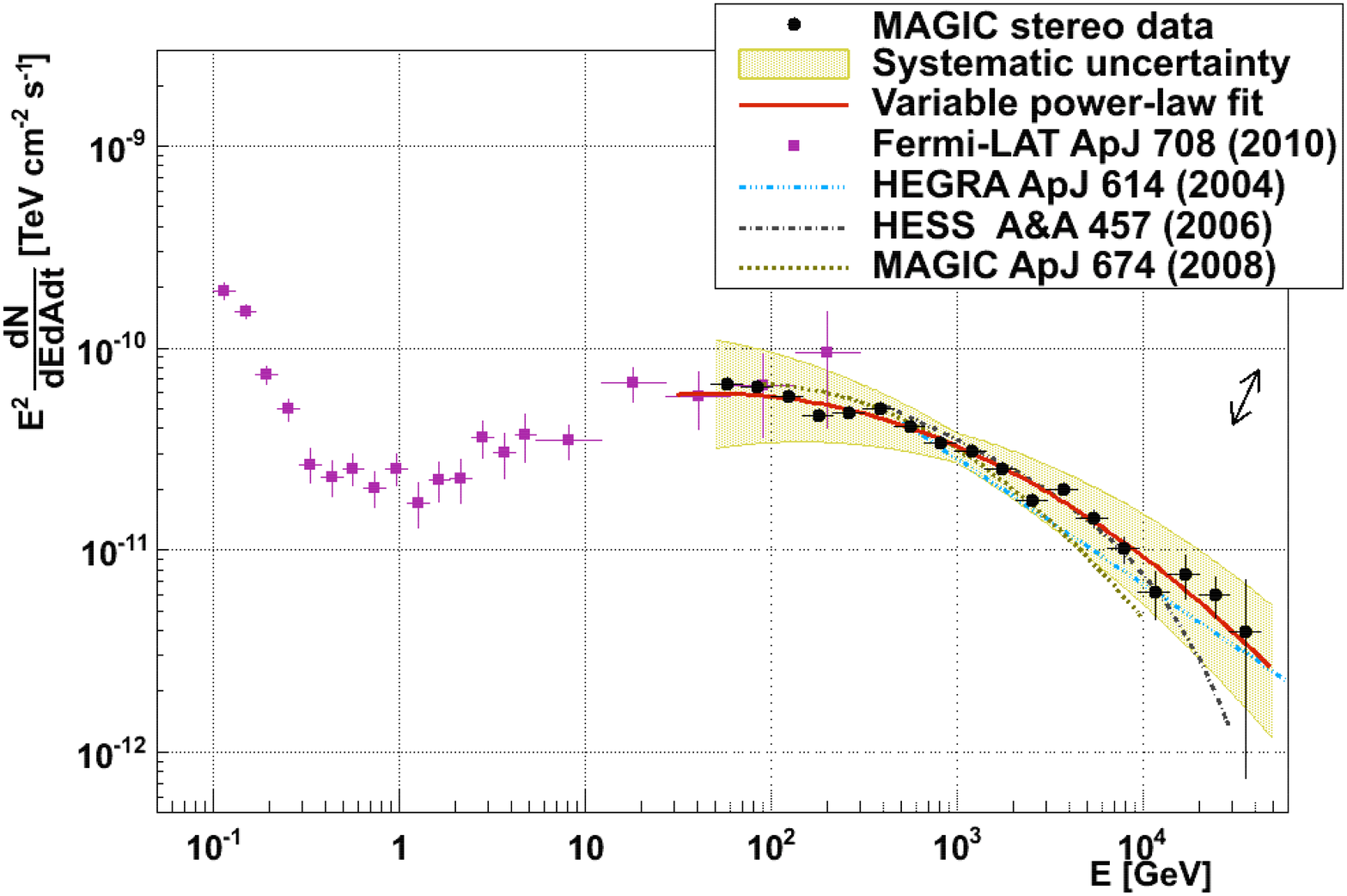}
  \caption{\small{
    Spectral Energy Distribution of the Crab Nebula obtained with the MAGIC telescopes, 
    together with previous results from IACTs. The black arrow indicates the systematic 
    uncertainty on the energy scale. Figure from \cite{icrc_Crab_nebula}.
  }}
  \label{fig:Crab_sed}%
\end{figure}

Fig. \ref{fig:Crab_sed} shows the combined fit of the Fermi/LAT\cite{fermi_Crab_nebula}
and MAGIC data to a curved power law, along with results from other IACTs.
The peak of the IC component of emission is found at 59$\pm$6 GeV (statistical errors only).
Unfortunately systematic uncertainties make impossible to draw any conclusions regarding the
cutoff at $>$~10TeV energies.

\begin{figure}
  \centering
  \includegraphics[width=\columnwidth]{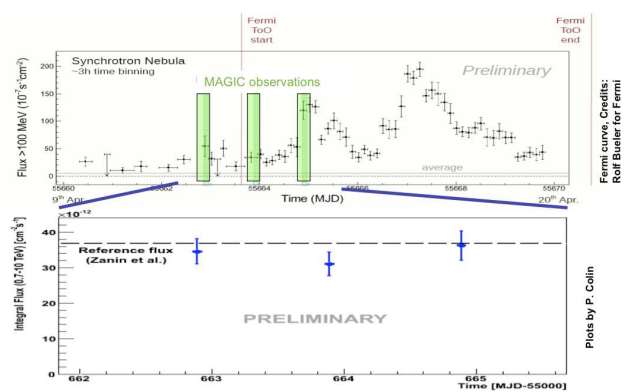}
  \caption{\small{
      Light curves of the Crab Nebula measured by Fermi/LAT at E$>$100 MeV and MAGIC
      in the range 0.7-10 TeV during a large GeV outburst on April 14th, 2011 (MJD 55665).
      There is no indication of variability at TeV energies.}}
  \label{fig:Crab_variability}%
\end{figure}

After recent reports of fast variability at GeV\cite{agile_flares,fermi_flares} energies and even 
TeV energies\cite{argo_flares}, MAGIC has searched for daily variations of the Crab Nebula flux
over the above-mentioned data sample, but also during a $\sim$3 hour observation 
period in April 2011 for which Fermi/LAT reports an increase of a factor 15 over the steady flux above
100~MeV\cite{fermi_flare_april2011}. MAGIC in fact observed also on the previous two nights. 
As shown in Fig. \ref{fig:Crab_variability} the flux for all three nights was found 
to be steady within statistical errors of $\sim$10\%.

\subsection{Crab Pulsar}

Models for high energy emission (polar cap, outer or slot gap) predict exponential or superexponential 
cutoffs in pulsar spectra at a few GeV and in fact that is what Fermi/LAT has been observing for
all pulsars at 100~MeV - 10~GeV (see e.g. \cite{fermi_pulsars}). MAGIC however observed pulsed emission
from the Crab pulsar above 25~GeV and a hint for emission above 60~GeV\cite{magic_Crab_science}.
Recently VERITAS reported detection of pulsed emission in excess of 100~GeV\cite{veritas_Crab_science}.

\begin{figure}
  \centering
  \includegraphics[width=\columnwidth]{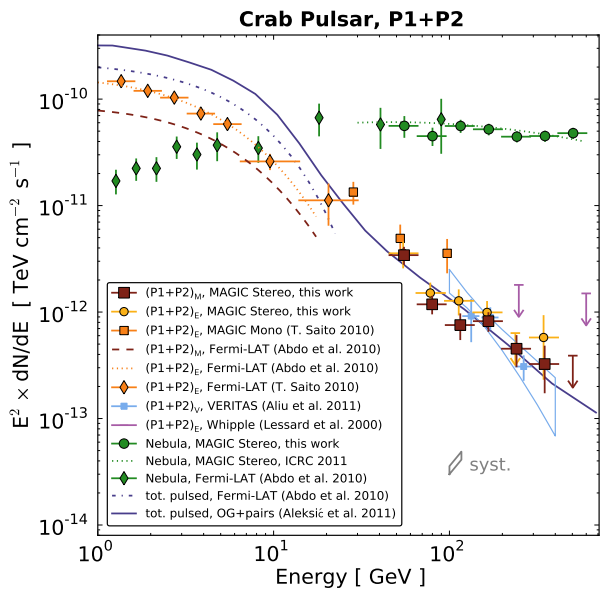}
  \caption{\small{
      Compilation of spectral measurements of MAGIC, VERITAS and
      Fermi-LAT for both peaks P1 and P2 together. Points of similar color refer
      to the same phase intervals (the indices of which refer to \textbf{M}AGIC,
      \textbf{E}GRET, \textbf{V}ERITAS definitions). See the original 
      paper\cite{stereo_Crab_pulsar} for a full discussion of the phase ranges
      We refer the reader to this paper also for the model displayed with a blue line: 
      in this model the spectral tails are inverse Compton radiation
      of secondary and tertiary electron pairs on magnetospheric IR-UV photons.
  }}
  \label{fig:Crab_pulsar}%
\end{figure}

An analysis of 59 hours taken from Oct.~2007 to Feb.~2009 with MAGIC-I alone resulted in separated 
spectra for the two peaks of pulsed emission P1 and P2, which show a power law shape between 25 GeV and 
100 GeV\cite{icrc_Crab_pulsar,mono_Crab_pulsar}. The measured flux of P1+P2 is 
signiﬁcantly higher than suggested by the extrapolation 
of the fitted cutoff function determined by Fermi-LAT (more than 5$\sigma$ even taking 
systematics into account)

\begin{figure}
  \centering
  \includegraphics[width=\columnwidth]{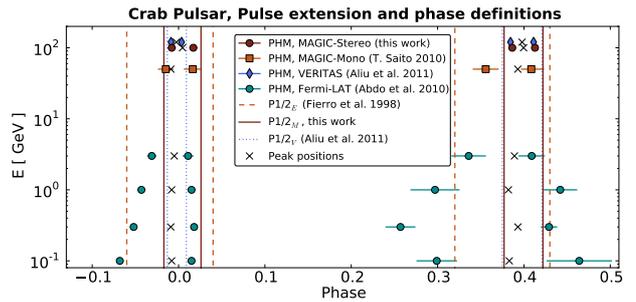}
  \caption{\small{
      Compilation of pulse profile parameters at different energies,
      measured by Fermi-LAT, MAGIC mono, MAGIC stereo
      and VERITAS. The crosses indicate the phase of the peak,
      while the colored points are the phases of the half-maxima (PHM). 
      Figure from \cite{stereo_Crab_pulsar}.
  }}
  \label{fig:Crab_peaks}%
\end{figure}

What is more, the analysis of 73 hours of new stereo data allows to detect pulsed emission up to 400 GeV.
Again the spectra for P1 and P2 can be measured separately. The spectra of both P1 and P2 are compatible with
power laws of photon indices $4.0\pm0.8$ (P1) and $3.42\pm0.26$ (P2), respectively, and the
ratio P1/P2 between the pulses is $0.54\pm0.12$. Fig. \ref{fig:Crab_pulsar} shows the 
combined spectrum of both peaks, for mono and stereo MAGIC data, and compares it to the spectra 
measured by Fermi/LAT\cite{fermi_Crab_nebula} and VERITAS. This actually constitutes the first $\gamma$-ray
spectrum of the Crab Pulsar from 100~MeV to 400~GeV with no gap. 
The individual spectra are in good agreement within systematics and extend well beyond 
the exponential/superexponential cutoff at an energy of a few GeV predicted by most of the
polar cap or outer gap models. The peaks are significantly
narrower than in the GeV regime, and along with MAGIC-I and VERITAS data, a consistent 
trend in the width of the peaks and in the amplitude ratio P1/P2 from GeV to beyond
100 GeV can be established (Fig.~\ref{fig:Crab_peaks}).

\subsection{SNR/PWN}

Stereo observations with the MAGIC system allow to extract higher-resolution sky maps down to energies 
around 100~GeV. Separate maps of the H.E.S.S. unidentified sources\cite{hess_scan} 
HESS~J1857+026\cite{icrc_1857} 
and HESS~J1923+141\cite{icrc_w51} could be produced at the energy bands 150-700 GeV 
and above 700~GeV. The first source may show morphological evolution with energy. Its general
morphology and spectral shape support a PWN origin for the $\gamma$-ray emission.

The second source is in the field of view of the SNR W51C and has been 
detected by Fermi/LAT at GeV energies\cite{fermi_w51}. The emission measured by
MAGIC is spatially coincident with that reported by Fermi/LAT, and the
measured SED is in agreement with the Fermi/LAT and H.E.S.S. measurements. 
MAGIC's higher angular resolution shows that the bulk of the VHE $\gamma$-ray emission
comes from the shocked molecular cloud located where the SNR shell
collides with a large molecular cloud observed in radio data. This fact, and the better agreement
of radio data with the hadronic scenario, suggests that the $\gamma$-ray
emission is most likely of hadronic origin.

\subsection{$\gamma$-ray Binaries}

In the past decade, IACTs have established that the luminosity of some binary systems 
is dominated by $\gamma$-rays. Hence they are refered to as ``$\gamma$-ray binaries''. 

A multiyear study \cite{lsi1,lsi2,lsi3,lsi4} has revealed that the $\gamma$-ray binary 
LS I+61~303 is a variable
and periodic source. The VHE emission decreased significantly in 2008, but, thanks to the increased
sensitivity of the stereo system, MAGIC detected it significantly\cite{icrc_lsi} at a 
flux of less than 5\% of Crab
during this state of low emission in 2009. In Fall 2010 to Spring 2011 LS I+61 303 had returned to its 
previous level of VHE emission in excess of 10\% of Crab. This may points to a correlation with
super-orbital variability\cite{gregory} due to modulation of the circumstellar wind.

MAGIC has also observed the source HESS J0632+057\cite{icrc_0632}, which had been studied by 
H.E.S.S.\cite{hess_0632} and VERITAS\cite{veritas_0632} and tentatively associated to a 
binary system with a period of 
$\sim$320 days in X-rays. VHE emission has indeed been detected in coincidence with the 
foreseen 2011 X-ray outburst.

Both of the aforementioned binaries may well be powered by the shock between a pulsar wind
and the companion stellar wind. There is no evidence so far of VHE emission from microquasar 
jets\cite{icrc_microquasars}. In this respect, MAGIC has recently completed a 4~year long 
multiwavelength campaign aimed at detecting Cyg X-3 at VHE energies\cite{cyg_x3}. No evidence for VHE 
emission has been found and stringent upper limits have been set for the whole data set, for
periodicity with the 4.8~hour orbital period, for emission in coincidence with
low-hard or high-soft X-ray states, shortly after radio flares or simultaneous with flares
at GeV energies. 

MAGIC monitored Cygnus X-1 for $\sim$50 hrs in 2006, and found an
evidence of signal at the level of 4.1$\sigma$ (post-trial) on
September 24th, 2006 \cite{cyg_x1}. The source was in the low/hard state and in
coincidence with a hard X-ray flare. Following this promising result, 
MAGIC observed the source for an additional 90 hours between July 2007 and
November 2009. The last campaign was performed with the
stereoscopic system. However, no significant VHE signal was
found in this sample. 

A search for VHE emission in Sco X-1 in the Horizontal Branch,
a state when the radio and hard X-ray fluxes are higher and a powerful relativistic jet is present,
has also proved unfruitful\cite{sco_x1}.

\section{Extragalactic observations}

IACTs have now established new classes of VHE active galaxies different 
from blazars. Two starburst galaxies have been discovered (NGC~253 by H.E.S.S.\cite{hess_ngc253} 
and M~82 by VERITAS\cite{veritas_m82}), 
representing a class of objects with no central source and whose $\gamma$-rays are probably
due to the global emission of cosmic rays in the galaxy. Besides, and among other non-blazar classes,
MAGIC has discovered two of the three quasars established at VHE and two of the four VHE radiogalaxies.
Blazars, however, still constitute the majority of extragalactic objects detected at this energy range.

\subsection{Radiogalaxies}

The study of radiogalaxies may bring new insights into the process accelerating particles
in jets. This is so because, in contrast to blazars, they are nearby, so we may be able to 
pinpoint the VHE emission region using input also from other wavelengths, and because their jets are
not aligned with the line of sight, so beaming effects are not so significant. It must be said, 
however, that blazar-like emission models have been proposed for radiogalaxies.
This class of objects may also be responsible for ultrahigh energy cosmic 
rays\cite{uhe_radiogalaxies}. 

MAGIC is so far responsible for detecting two out of the four VHE radiogalaxies, namely
NGC~1275\cite{icrc_ngc1275} and IC~310\cite{ic310}, both in the Perseus cluster. NGC~1275 was detected 
during a period of enhanced GeV activity reported by Fermi/LAT\cite{1275_fermi_flare}. 
Its spectrum is very steep,
with a photon index -4.0$\pm$0.4. The detection was in fact only possible thanks to the improved
sensitivity at energies below 100~GeV brought about by the stereo system.
IC~310 is the only radiogalaxy of the ``head-tail'' type detected so far. In contrast to 
NGC~1275, it displays a very hard spectrum with photon index 2.00$\pm$0.14. 

M~87 was the first radiogalaxy discovered at VHE (by HEGRA\cite{m87_hegra} ) 
and has been extensively studied by H.E.S.S., 
VERITAS and MAGIC. It is nearby (17~Mpc) and well characterized at other wavelengths.
Observations in 2008, using these three IACTs and radio VLBI\cite{m87_science}, suggested that
VHE $\gamma$-rays originated very close to the central black hole (a few Schwarzschild radii, 
with a Schwarzschild radius of $\sim$100 A.U.). 

\begin{figure}[!t]
  \vspace{5mm}
  \centering
  \includegraphics[width=0.95\columnwidth]{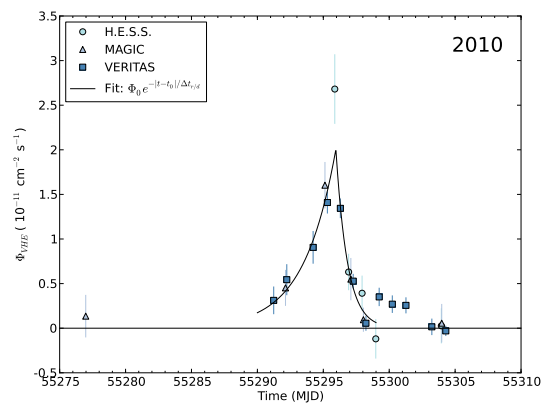}
  \caption{\small{VHE light curve of M 87 during the 2010 ﬂare. The line represents the 
      exponential fits to the data, which result in a rise time of 1.7 days and a decay time of 0.6 days. 
      Figure from \cite{m87_mw}.}}
  \label{fig:m87}
\end{figure}

Further cooperation of the three instruments and more multiwavelength observations\cite{m87_mw} 
added essential information to the understanding of M~87, but may indeed have complicated this interpretation. 
In total, three flares have been recorded at VHE: one in 2005, one in 2008 and one in 2010. They
show significant differences:
\begin{itemize}
\item In 2010 the coordinated observations led to the best-sampled VHE light
curve during a flaring state from this source (21 observations in 15 days, see Fig. \ref{fig:m87}), 
revealing a single, isolated outburst. The
measured VHE light curve of the flare is well described by an exponential rise and decay,
while the other two VHE flares do not show a clear time structure.
\item VLBA 43~GHz observations, triggered by the detection of
the VHE ﬂare, show no indications for an enhanced radio
emission from the jet base in 2010. This is in contrast to the above-mentioned 
observations in 2008.
\item In 2008 and 2010 the VHE flares are accompanied by a high state of the X-ray core with a flux 
increase by a factor $\sim$2, while in 2005 the strong flux dominance (more than
a factor 30) of the nearby X-ray feature HST-1 could have suppressed the detection of such an 
increase of the core emission. 
\end{itemize}
It is thus unclear how the $\gamma$-rays were produced and one may actually wonder if all flares came 
from the same emission site and are produced through the same mechanism. 
It must be said however that all flares share very similar time-scales (around 1 day), 
peak fluxes (F($>$0.35 TeV) = (1-3)$\times$10$^{−11}$ ph cm$^{-2}$ s$^{-1}$ ), and 
VHE spectra. In fact, MAGIC has recently studied this object during its low 
state\cite{icrc_m87,m87_low} and found a spectral index 
which is consistent with that measured during the high state. This indicates that the same emission 
mechanism may be responsible for both the high and low states of emission.

\subsection{Quasars}

MAGIC has claimed the detection of two out of three quasars known to emit at VHE. Quasars are distant
objects and, as such, instrumental to constrain the level of Extragalactic Background Light (EBL) which
attenuates the flux of $\gamma$-rays of GeV to TeV~energies. Besides, the intense optical/UV
radiation field present
in quasars results in strong attenuation intrinsic to the source and is expected to distort the 
VHE spectrum. 

\begin{figure}[!t]
  \vspace{5mm}
  \centering
  \includegraphics[width=\columnwidth]{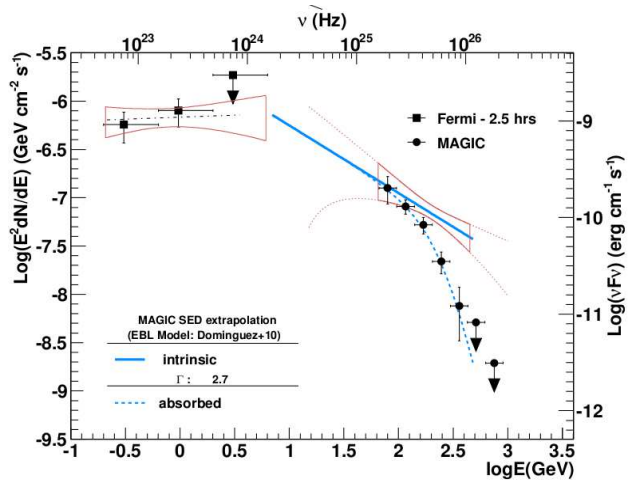}
  \caption{\small{High energy SED of PKS~1222+21 during the flare of 2010 June 17 (MJD 55364.9), 
      showing Fermi/LAT (squares) and MAGIC (circles) differential fluxes. 
      A red bow tie in the MeV/GeV range represents the uncertainty of the likelihood fit to the 
      Fermi/LAT data.
      The unfolded and deabsorbed spectral fit of the MAGIC data is also shown as a grey bow tie, 
      extrapolated to lower and higher energies (dotted lines). A thick solid line (photon index $\Gamma=2.7$) 
      indicates a possible extrapolation of the MAGIC deabsorbed data to lower energies. 
      The thick dashed line represents the EBL absorbed spectrum obtained from the extrapolated 
      intrinsic spectrum using  the model of \cite{Dominguez2010}. Figure from \cite{icrc_quasars}.
    }}
  \label{fig:pks1222}
\end{figure}

3C~279 (z=0.536) is the farthest VHE source detected so far. It was discovered by MAGIC at VHE in 
2006\cite{3c279_science} and allowed to place very strong constraints on the density of 
EBL, which have only now been confirmed by the detection of a second quasar. 

This second quasar, 
PKS~1222+21 (4C~+21.35, z=0.432), has been discovered by MAGIC in 2010\cite{pks1222,icrc_quasars}. 
The detection of PKS~1222+21 coincides with high GeV $\gamma$-ray activity
measured by Fermi/LAT\cite{fermi_pks1222}. 
The VHE flux varies significantly within the 30~minutes of exposure 
implying a flux doubling time of about 10~minutes. 
As shown in Fig. \ref{fig:pks1222} the
VHE to GeV spectrum, corrected for the absorption by the EBL, can be described by a single power 
law with photon index $2.72\pm0.34$ between 3~GeV and 400~GeV, consistent with $\gamma$-ray emission 
belonging to a single component in the jet. 
The absence of a spectral cutoff at 30-60~GeV (indeed, one finds a strict lower
limit $E_c>$~130~GeV) constrains the $\gamma$-ray emission region to lie outside the
broad line region, which would otherwise absorb the VHE $\gamma$-rays. Together with the
detected fast variability, this challenges present emission models from jets in FSRQs.

Recent MAGIC observations of 3C~279\cite{3c279,icrc_extragalactic} show the same apparent 
contradiction. Some explanations have already been proposed in order to solve it, 
invoking the presence of very compact emission regions embedded within the large 
scale jet or the possibility of a very strong jet 
recollimation (see \cite{icrc_quasars} for more details).

\subsection{Blazars}

Blazars still constitute the vast majority of extragalactic sources detected at VHE. About 40
blazars have been detected at this energy range, 14 of them by MAGIC.

\begin{figure}[!t]
  \vspace{5mm}
  \centering
  \includegraphics[width=\columnwidth]{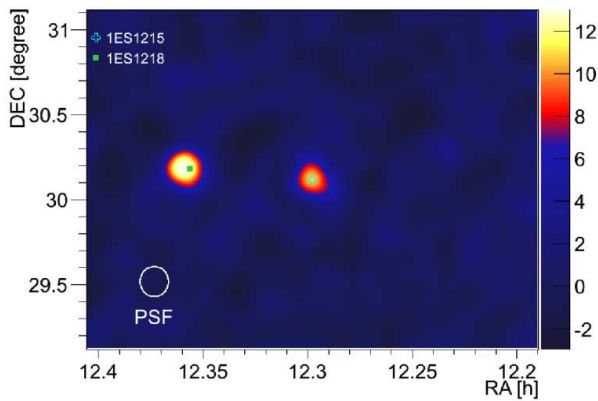}
  \caption{\small{
      Significance skymaps corresponding to the 2011 observations of the field of view of
      1ES~1215+303 and 1ES~1218+304, at a time when both objects were active in VHE
      $\gamma$-rays. Figure from \cite{icrc_1215_1218}.
    }}
  \label{fig:1215_1218}
\end{figure}

Among the most recent results regarding this class of objects are:
\begin{itemize}
\item The detection of 3C~66A\cite{icrc_3c66a} in December 2009 and January 2010 with a 
photon index 3.64$\pm$0.39(stat)$\pm$025(sys). From the spectra corrected for absorption by the 
EBL the redshift of the blazar can be constrained to z$<$0.68. 
\item First multiwavelength campaign including VHE coverage on the HBL 1ES~1011+496\cite{icrc_1011}. 
The campaign resulted in the first confirmation of the source at VHE at a flux level 
similar to that measured during the discovery. No apparent correlation between the 
flux levels in VHE, optical and X-rays was found. The data was modelled by a one-zone 
SSC and a self-consistent two-zone SSC model 
which fit the points well indicating that the SSC scenario is sufficient to describe 
the emission mechanisms of the source.
\item Observations of the two BL Lac objects 1ES~1215+303 and 1ES~1218+304\cite{icrc_1215_1218}, 
in the same field of view for MAGIC and separated by 0.8 deg, in 2010 and 2011 (see Fig. 
\ref{fig:1215_1218}). It resulted in the first detection at VHE of 1ES~1215+303 and 
a high quality measurement of the spectrum of 1ES~1218+304, 
an already known VHE gamma-ray emitter\cite{1218}.
\item Dedicated monitoring observations of the nearby BL Lacs Mrk~501, Mrk~421, 1ES~1959+650 since 
2006\cite{icrc_monitoring},
meant to estimate their duty cycle, investigate potential spectral changes during periods of 
different source activity, correlate the results with multiwavelength observations and search for 
potential periodic behavior.
\item Two multiwavelength campaigns\cite{icrc_180_2344} on the high-frequency peaked blazars Mrk~180 and 
1ES~2344+514 involving instruments from radio to VHE (RATAN-600, Mets\"ahovi, Effelsberg, VLBA, IRAM, 
KVA, Swift, AGILE, Fermi/LAT and MAGIC-I). 
\item Very recent discovery of a new blazar, 1ES~1741+196, whose host of the 
largest and most luminous in all BL Lacs, and MG4~J2001+438\cite{c23_taup}. We refer the reader to 
\cite{icrc_extragalactic} for more details and reports on other extragalactic observations.
\end{itemize}

In general, optical monitoring of a selected number of targets with the KVA optical telescope 
has proved to be a successful means to discover new AGNs at VHE. KVA observes concurrently
with MAGIC in La Palma and triggers VHE observations when a source reaches 
a high optical state. During mono operation with MAGIC-I, this strategy 
resulted in the discovery of Mrk~180 (HBL, z=0.045), 1ES~1011+496 (HBL, z=0.212) and
S5~0716+714 (LBL, z$\sim$0.31). During the last two years of stereo observations,
in the discovery of B3~2247+381 (HBL, z=0.119) and 1ES~1215+303 (HBL, z=0.130 or z=0.237).

\begin{figure}[!t]
  \vspace{5mm}
  \centering
  \includegraphics[width=\columnwidth]{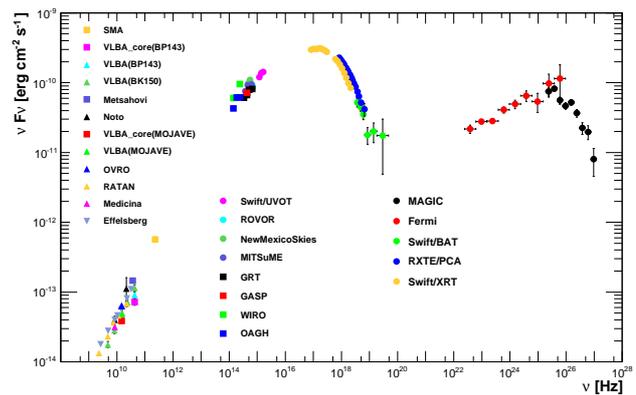}
  \caption{\small{
      Spectral energy distribution of Mrk~421 averaged over all
      the observations taken during the multifrequency campaign from
      2009 January 19 (MJD 54850) to 2009 June 1 (MJD 54983). The legend reports the 
      correspondence between the instruments and the measured fluxes. 
      The reader may find more details and the original figure in \cite{421}.
    }}
  \label{fig:mrk421}
\end{figure}

During the past years, MAGIC has organized a good number of multiwavelength observations 
involving many instruments, lately also Fermi/LAT. Strictly simultaneous broad-band 
SEDs are key to modeling blazars in a reliable manner. Besides the above-mentioned examples, this
is well illustrated by recent simultaneous observations 
of Mrk~421 in \cite{421,icrc_421_501,421_2}, which resulted in the
SED shown in Fig. \ref{fig:mrk421}. The unprecedented sampling allows to put tight 
constraints on the model parameter space.

\subsection{Gamma Ray Bursts}

MAGIC was especially designed to search for the prompt emission of GRBs. The telescope can move 
180 deg in 20 seconds and does so in a totally automatic manner after an alert from GCN.
MAGIC is observing an average of about one GRB every month. No detection has been achieved 
so far. MAGIC has lately reported a limit on the afterglow emission\cite{grb_afterglow}.

In Spring 2011 MAGIC observed GRB~110328 (Swift~J164449.3+573451), later disregarded as GRB due to 
long-lasting activity and which probably represents the birth of a black-hole relativistic jet
\cite{strange_grb1,strange_grb2}. MAGIC observed starting 2.5 days 
after the onset of the event detecting no significant excess. Preliminary upper limits have 
been extracted\cite{icrc_extragalactic}.

\section{Cosmic ray measurements, galaxy clusters and dark matter searches}

\subsection{Cosmic ray measurements}

\begin{figure}[!t]
  \vspace{5mm}
  \centering
  \includegraphics[width=\columnwidth]{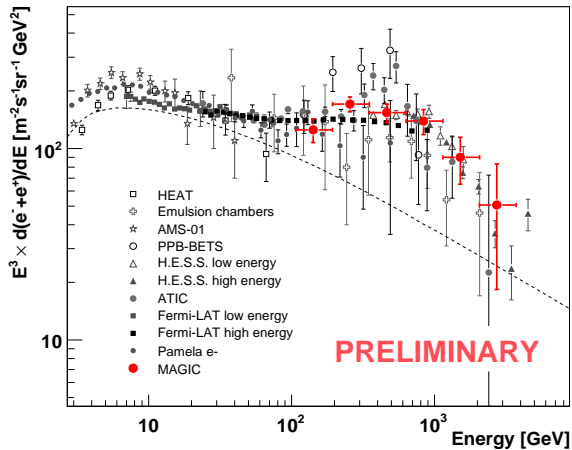}
  \caption{\small{
      Spectrum of electrons and positrons measured with MAGIC, along with other
      measurements from instruments on orbit, balloons and IACTs. MAGIC extends 
      the spectrum produced by IACTs down to 100~GeV.
      Figure from \cite{icrc_electrons}.
    }}
  \label{fig:electrons}
\end{figure}

IACTs may discriminate showers initiated by electrons or positrons from the 
background of hadronic cosmic ray showers through the image shape. A first sample of 14 hours 
of extragalactic observations has allowed to measure the combined e$^-$/e$^+$ spectrum in the 
energy range between 100~GeV and 3~TeV\cite{icrc_electrons}.
The preliminary result is shown in figure \ref{fig:electrons}. The spectrum
can be fitted to a power-law with differential index 
3.16$\pm$0.06(stat)$\pm$0.15(sys). The spectrum is in good agreement with previous measurements 
(the bump observed by ATIC cannot be excluded or confirmed though).

The fact that electrons and positrons deflect in opposite directions in the geomagnetic
field means that the shadows cast by the Moon in the distribution of arrival directions
of electrons and positrons do not coincide. By pointing the telescopes to these two positions
and discriminating electron/positron events from the hadronic background, one can measure the
ratio of electrons to positrons at the energy range from hundreds of GeV to some TeV. 
Even if the technique is challenging because of the high background light induced by the 
Moon, the MAGIC collaboration has managed to performed for the first time such observations 
in 2010 and 2011\cite{icrc_moon_shadow}. The detection of the electron Moon shadow will however require
several years of observations due to the short observation time which is available every year
at the right phase of the Moon and zenith angle.

\subsection{$\gamma$-rays from cosmic rays and dark matter annihilation in galaxy clusters}

Cluster of galaxies are expected to be significant $\gamma$-ray emitters for several reasons.
To begin with, they are still actively evolving objects, so they should dissipate energies of the 
order of the final gas binding energy through merger and accretion shocks as well as turbulences,
all of which may accelerate electrons and protons to VHE. Besides clusters host powerful
sources like radio galaxies and supernova-driven galactic winds. In addition, clusters 
often show direct evidence for shocks and turbulence as well as relativistic particles. To
conclude, galaxy clusters present very large mass--to--light ratios (M/L) and considerable Dark Matter 
(DM) overdensities, which make them into good targets to search to $\gamma$-rays produced in the
annihilation or decay of DM particles.

The Perseus cluster, at a distance of 78~Mpc, is the brightest X-ray cluster and
hosts a massive cooling flow and a luminous radio mini-halo that fills a large fraction of the 
core region. MAGIC conducted the deepest survey ever made at VHE of the Perseus cluster, collecting data 
both in mono mode ($\sim$25~hours between November and December 2008 
\cite{perseus_mono}) and in stereoscopic mode ($\sim$90~hr between October 2009 and 
February 2011\cite{icrc_perseus}).

These observations already resulted in the aforementioned detection of the two radiogalaxies NGC~1275
and IC~310. Besides, mono observations permitted to constrain the average CR-to-thermal 
pressure to $<4\%$ for the cluster core region and to $<8\%$ for the entire cluster. Stereo observations 
allow to significantly tighten the previous constraints and to start to probe the acceleration 
physics of CRs at structure formation shocks\cite{perseus_mono_predictions}. The estimation 
and interpretation of the flux upper limits, however, are still ongoing.

Mono observations were also used to limit the emission produced by DM annihilation\cite{perseus_mono}. 
The cluster DM density profile was modeled with a Navarro-Frank-White profile 
and the DM particle was assumed to be the neutralino within the mSUGRA scenario.
Even considering one of the most optimistic values for the particle physics factor, the
predicted fluxes are still of the order of $10^4$ times below the measured upper limits.
Nevertheless it is worth mentioning that Perseus has turned out to be a challenging
target for DM searches, precisely because of the presence of bright $\gamma$-ray sources
in the field of view, and because the expected flux is much smaller than that coming from 
CR interactions and may be too extended (see \cite{icrc_perseus} and references within).

\subsection{Other DM searches}

Along with the observations of the Perseus cluster, MAGIC is searching for emission
produced by DM annihilation in so-called ``Unassociated Fermi Objects'' (UFOs). These 
are sources detected by Fermi/LAT\cite{1_fermi_cat} which do not have a counterpart at other wavelengths.
UFOs far from the galactic plane and displaying especially hard spectra may be associated
to mini-halos of DM in our galaxy\cite{icrc_clumps}. So far, MAGIC has produced upper limits to
the emission of two UFOs, 1FGL~J0338.8+1313 and 1FGL~J2347.3+0710, after relatively short
exposures of the order of 10 hours.

Dwarf spheroidals (dSph) orbitting our galaxy are prime candidates to search for DM annihilation
signals on account of their high M/L and their short distances. MAGIC has observed 
the dSph Draco\cite{draco}, Willman-1\cite{willman} and lately Segue-1\cite{segue,icrc_segue}.
Segue1 is considered by many as the most DM dominated satellite galaxy near 
our galaxy. The 30 hour MAGIC observation represents the largest survey ever made 
on a single dSph by IACTs. No significant gamma-ray emission was found. A novel
analysis takes into account the spectral features of the $\gamma$-ray spectrum 
of specific DM models in a Supersymmetric scenario\cite{icrc_segue}. Limits are
however still a factor 600 above the most optimistic expectations of these models.
More observations of dSph are under way.

\section{MAGIC Upgrade}

Besides field tests of novel types of photodetectors which may significantly boost the
photon detection performance of MAGIC\cite{icrc_HPD}, the most significant activity at the
MAGIC site is a general upgrade scheduled this year and in 2012. 

The MAGIC telescopes were 
shut down on June 15th, 2011, to perform several major hardware inverventions:
\begin{itemize}
\item Both telescopes will be equipped with a new 2 GSamples/s readout based on DRS4 chip 
(linear, low dead time, low noise).
\item The camera of MAGIC-I will be upgraded from 577 to 1039 pixels to match the camera geometry 
and the trigger area of MAGIC-II.
\item Both telescopes will be equipped with ``sumtrigger'' covering the total conventional 
trigger area\cite{icrc_sumtrigger}. 
\end{itemize}

After the upgrade, we expect some improvement in sensitivity for point sources and significantly better 
performance for extended sources. The introduction of sumtriggers in both telescopes will lead to
a reduction of the energy threshold. The intervention also aims at making the hardware of both 
telescopes essentially equal, thereby making the maintenance and operation easier in the future.

\section{Conclusions}

The MAGIC two 17 meter diameter Very High Energy (VHE) $\gamma$-ray telescopes have now operated 
for two years in stereoscopic mode. Based on a long Crab nebula observation in the last two
winters, the performance of the instrument has been evaluated: the analysis threshold 
energy has decreased to 50 GeV, the integral sensitivity at its optimal range above 300~GeV 
is 0.76\% Crab units, the differential sensitivity 10\% Crab units at sub-100~GeV energies. 

Highlights of the last two years of observations are the
measurement of the Crab Nebula spectrum over three decades in energy, from $\sim$50~GeV 
to $\sim$50~TeV, which has made a precise determination of the position of the IC peak
possible; the detection of the Crab pulsar up to an energy of 400~GeV, with energy spectra measured 
for both P1 and P2 and no evidence for a spectral cutoff; the discovery of two new 
radiogalaxies at VHE (NGC~1275 and IC-310) displaying very different spectra; the absence of 
an energy cutoff and the discovery of fast variability in the quasars 3C~279 and PKS~1222+21, both facts
challenging current models of VHE emission in quasars; the discovery at VHE and the characterization 
of numerous blazars; upper limits to the VHE emission of the Perseus cluster of galaxies and 
to Dark Matter annihilation in unasociated Fermi objects and dwarf Spheroidals; and 
the measurement of the electron+positron spectrum between 100~GeV and 3~TeV. 

A major upgrade of the readout and trigger electronics of both telescopes, and of the camera of the 
first MAGIC telescope are currently ongoing.

\section*{Acknowledgments}

The author would like to thank the organizers of the 32nd ICRC for the opportunity to 
present the latest results of MAGIC at the conference and his colleagues in the collaboration
for their numerous comments and corrections. 
The MAGIC collaboration also thanks the Instituto de Astrof\'{\i}sica de Canarias for
the excellent working conditions at the Observatorio del Roque de los Muchachos
in La Palma. The support of the German BMBF and MPG, the Italian INFN, the
Swiss National Fund SNF, and the Spanish MICINN is gratefully acknowledged. This
work was also supported by the Marie Curie program, by the CPAN CSD2007-00042
and MultiDark CSD2009-00064 projects of the Spanish Consolider-Ingenio 2010
programme, by grant DO02-353 of the Bulgarian NSF, by grant 127740 of the
Academy of Finland, by the YIP of the Helmholtz Gemeinschaft, by the DFG Cluster
of Excellence ``Origin and Structure of the Universe'', and by the Polish MNiSzW
grant 745/N-HESS-MAGIC/2010/0.


\clearpage

\end{document}